%% file: led.tex
\documentclass[aps,prl,twocolumn,showpacs,groupedaddress]{revtex4}  

\usepackage{d0_style}
\usepackage{graphicx}  
\usepackage{dcolumn}   
\usepackage{bm}        
\usepackage{amssymb}   

\begin{document}

\hspace{5.2in} \mbox{Fermilab-Pub-08/061-E}

\title{Search for Large Extra Dimensions via \\
Single Photon plus Missing Energy Final States at $\sqrt s = 1.96\ {\rm TeV}$ }
\input list_of_authors_r2.tex  
\date{June 30, 2008}

\begin{abstract}
We report on a search for large extra dimensions in a data sample of 
approximately $1\ {\rm fb^{-1}}$ 
of \ppbar~collisions at $\sqrt s  = 1.96\ {\rm TeV}$.  We investigate
Kaluza-Klein graviton production with a photon and missing transverse 
energy
in the final state.  
At the 95\% C.L. we set limits on the fundamental
mass scale $M_{D}$ from $884\ {\rm \gev}$ to $778\ {\rm \gev}$ for
two to eight extra dimensions.
\end{abstract}

\maketitle 


Arkani-Hamed, Dimopoulos, and Dvali (ADD)~\cite{add} 
made the first attempt to solve
the hierarchy problem of the standard model (SM)
by postulating the existence of $n$ new large extra
spatial
dimensions (LED).  
In this approach, 
the SM particles are confined
to a 3-dimensional brane while gravity is diluted
in the larger volume.  
The size of the compactified extra space ($R$), the effective
Planck scale in the $4$-dimensional
space-time ($M_{Pl}$), and the fundamental 
Planck scale in the $(4+n)$-dimensional
space-time ($M_{D}$), are related by the expression
$M_{Pl}^2 = 8\pi M_{D}^{n+2}R^{n}$.
Due to the compactification of the extra space,
the gravitational field appears as a series of 
quantized energy states,
which are referred to as Kaluza-Klein modes. 
A Kaluza-Klein graviton ($G_{KK}$) behaves
like a massive, non-interacting, stable particle
whose direct production gives an 
imbalance in the final state momentum
as its collider signature.  

In this Letter we report 
the results of a search for
LED in the final state with a single 
photon plus missing transverse energy ($\gamma+\met$), 
using data collected with the D0 detector at the Fermilab 
Tevatron collider.
This signature arises from the process
$q\bar{q}\rightarrow\gamma G_{KK}$, which is 
studied in detail in~\cite{giudice}.
The CDF collaboration
carried out a similar search with 87~\ipb~of data, 
setting 95\% C.L. lower limits on $M_{D}$ 
of 549, 581, and 601 \gev~ for 4, 6, and 8
extra dimensions, respectively~\cite{cdfrun1}.
Searches for LED in other final states have been performed
by collaborations at the Tevatron~\cite{tevatron,cdfmonojet} 
and the CERN LEP collider~\cite{lep1}.

The background to the $\gamma+\met$ signal is dominated by
electroweak boson production and non-collision
background where muons from the beam halo or cosmic
rays undergo bremsstrahlung and produce an energetic photon.  The
electroweak background is dominated by the processes
$Z+\gamma \rightarrow \nu\overline{\nu}+\gamma$,
$W\rightarrow e\nu$ where the electron
is misidentified as a photon, $W+\gamma$ where the
lepton from the $W$ boson decay is not detected,
and $W/Z + \text{jet}$ production 
where the jet is misidentified 
as a photon.

The D0 detector~\cite{nim}
comprises a central-tracking
system with a silicon microstrip tracker (SMT) 
and a central fiber tracker (CFT), 
both housed within a $2\ {\rm T}$ superconducting solenoidal 
magnet, with designs optimized for tracking and 
vertexing at $|\eta|<3$ and $|\eta|<2.5$, respectively, where
$\eta$ is the pseudorapidity~\cite{pseudo} measured
with respect to the geometrical center of the detector.
The central preshower system (CPS) is located
in front of a liquid-argon/uranium 
calorimeter and consists of three
layers of scintillating
strips, providing precise measurement of electromagnetic (EM) shower positions.
The calorimeter has a central 
section (CC) covering $|\eta| \leq 1.1$, and two end
sections (EC) that extend coverage to $|\eta|\approx 4.2$~\cite{run1det}.  Each
part contains an EM section closest to the
interaction region followed by fine and coarse hadronic sections.
The EM section has four longitudinal
layers and transverse segmentation of $0.1\times 0.1$ in
$\eta - \phi$ space (where $\phi$ is the azimuthal angle), with
the exception of the third layer, where it is $0.05\times0.05$.
Additionally, scintillators between the CC and EC cryostats provide 
sampling of developing showers for $1.1<|\eta|<1.4$.
The outer muon system, covering $|\eta|<2$, 
consists of a layer of tracking detectors and scintillation trigger 
counters in front of $1.8\ {\rm T}$ iron 
toroids, followed by two similar layers 
after the toroids.
The data in this analysis were recorded
using triggers requiring 
at least one energy cluster in the 
EM section of the calorimeter with
transverse momentum $p_{T}>20\ {\rm GeV}$. The triggers
are almost $100\%$ efficient
to select signal events. This set of data corresponds 
to an integrated luminosity of $1.05 \pm 0.06\ {\rm fb^{-1}}$~\cite{lumi}. 

We identify a reconstructed calorimeter cluster as a photon
when it satisfies the following requirements: 
(i) at least $90\%$ of 
the energy is deposited in the EM section
of the calorimeter; (ii) the calorimeter isolation variable
${\cal{I}} = [E_{\text{tot}}(0.4)-E_{\text{em}}(0.2)]/E_{\text{em}}(0.2)$ is less than 0.07, 
where $E_{\text{tot}}(0.4)$ denotes the total energy deposited
in the calorimeter in a cone of radius
${\cal{R}}=\sqrt{(\Delta\eta)^{2}+(\Delta\phi)^{2}} = 0.4$, 
and $E_{\text{em}}(0.2)$
is the EM energy in a cone of radius ${\cal{R}}=0.2$; 
(iii) the track isolation variable, defined as the scalar sum 
of the transverse momenta of
all tracks that originate from the interaction vertex in an annulus
of $0.05<{\cal{R}}<0.4$ around the cluster, is less than $2\ {\rm GeV}$;
(iv) it has $|\eta|<1.1$;
(v) both transverse and longitudinal shower shapes are consistent
with those of a photon;
(vi) it has neither an associated track in the central tracking
system nor a significant density
of hits in the SMT and CFT systems 
consistent with the presence of a track with $p_{T}$
in agreement with its transverse energy; 
and (vii) there is an energy deposit in the CPS 
matched to it.
Jets are reconstructed using the iterative
midpoint cone algorithm~\cite{jet} with a cone size of $0.5$.
The missing transverse energy is computed from calorimeter
cells with $|\eta|<4$ and corrected for the EM and jet energy scales.

The {\it photon} sample is obtained by 
selecting events with only one photon with $p_{T}>90\ {\rm GeV}$,
at least
one reconstructed interaction
vertex consistent with the measured direction of
the photon (see below),
and $\met >70\ {\rm GeV}$.
Additionally, in order to avoid large $\met$ due to
mismeasurement of jet energy, we require 
no jets with $p_{T}>15\ {\rm GeV}$.  
The reduction of the signal efficiency due to the jet
veto on initial state radiation has been estimated
using {\sc pythia}~\cite{pythia} to be about $9\%$.
The applied \met~requirement guarantees 
negligible multijet background in 
the final candidate sample while being
almost fully efficient for signal selection.

We reject events with reconstructed muons 
and with cosmic ray muons identified using the timing
of the signal in the muon 
scintillation counters
or by the presence of a characteristic pattern of hits
in the muon drift chambers that is aligned with the reconstructed photon.
In order to further 
reject events with leptons that leave a distinguishable 
signature in the tracker but that are not reconstructed in the other
subsystems of the detector, 
we impose a requirement
on the $p_{T}$ of any isolated track 
not to be greater than $6.5$ \gev. 
A track is considered to be isolated
if the ratio between the scalar sum 
of the transverse momenta of
all tracks that originate from the interaction vertex in an annulus
of $0.1<{\cal{R}}<0.4$ around the track and the $p_{T}$ of the
track is less than $0.3$.

The EM pointing algorithm allows calculation
of the direction of the EM shower based on the 
transverse and longitudinal segmentation of the
calorimeter and preshower systems.  EM pointing is
performed independently in the azimuthal and polar
planes.  The former results in the measurement of the distance of closest
approach (DCA) to the $z$ axis (along the beam line), 
and the latter in the prediction
of the $z$ position of the interaction vertex in the
event, both with a resolution of about $2\ {\rm cm}$.  
We require that the $z$ coordinate of at least one interaction vertex in the
event be within $10\ {\rm cm}$ of the position predicted by the
pointing algorithm and use the DCA to estimate the remaining
background from jet-photon misidentification and non-collision events.
Misidentified jets
have poor pointing resolution, and therefore 
a wider DCA distribution compared to electrons or photons.  
Likewise, one can anticipate 
the DCA distribution for photon candidates in non-collision 
events to have an even wider shape.
After these requirements, 
$35$ events are selected in the {\it photon} sample.

We prepare three DCA distribution templates: the {\it non-collision}
template, the {\it misidentified jets} template, and the {\it $e/\gamma$} 
template.  
The first 
template is obtained from a sample in which a photon candidate, passing the
same quality requirements as for the {\it photon} sample, is selected 
from events with no
hard scatter (no
reconstructed interaction vertex or fewer than three reconstructed
tracks), or from events with identified cosmic muons.
The {\it misidentified jets} template is extracted from the
{\it fake photon} sample, which fulfills exactly the same requirements
 as the
{\it photon} sample except that the photon track isolation 
requirement is inverted.  This sample is dominated by
misidentified jets.
Finally, the {\it $e/\gamma$} template is obtained from a 
data sample of
isolated electrons. 

\begin{figure}[t]
\includegraphics[scale=0.44]{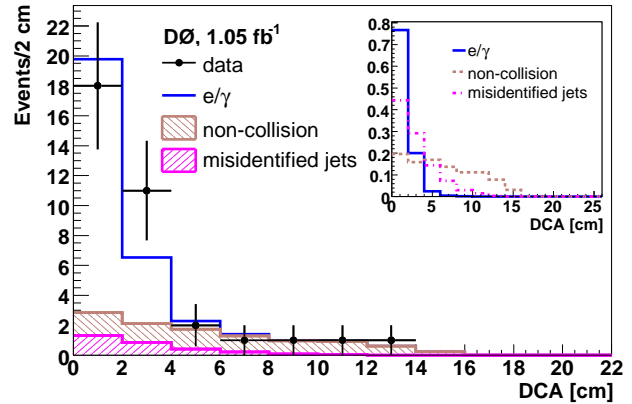}
\caption{\label{fig:shapes} DCA distribution for the selected events
in data (points with statistical uncertainties). The different
histograms represent the estimated background composition from the template
fit to this distribution. The inset figure compares the individual
template shapes.}
\end{figure}

The total number of background events
from misidentified jets ($N_{\text{misid}}$) can
be predicted from the {\it fake photon} sample based on the rates
at which jets, passing all other photon identification criteria, fail
or pass the track isolation requirement.  
To measure those rates we use an {\it EM plus
jet} sample, where the EM object passes all photon identification 
requirements except
the track isolation, and where the jet approximately balances
the EM object in the transverse plane.  We first determine
the number of events ($N_{1}$) in the sample that fail the track
isolation requirement.  
We then fit the DCA distribution of the events that pass 
the track
isolation to a linear sum of the {\it $e/\gamma$} and
{\it misidentified jets} templates in order
to extract the number of misidentified jets ($N_{2}$)
passing the track isolation.
$N_{\text{misid}}$ is then equal to the number of
events in the {\it fake photon} sample multiplied by $N_{2}/N_{1}$.
We fit the DCA distribution in the {\it photon} sample 
to a linear sum of the three templates, fixing
the contribution of {\it misidentified jets} as described
above, and determine the {\it $e/\gamma$} and {\it non-collision}
contributions.  The result of the fit is illustrated in 
Fig.~\ref{fig:shapes}.  
Most of the signal photons have DCA less than $4\ {\rm cm}$, 
therefore we limit our analysis to this particular
window, which contains $29$ data events.

The only physics background to the $\gamma+\met$ final
state is the process 
$Z+\gamma \rightarrow \nu\overline{\nu}+\gamma$.
This irreducible contribution is estimated from a 
sample of Monte Carlo (MC) events
generated with {\sc pythia} using
CTEQ6L1 parton distribution functions (PDFs)~\cite{cteq}.
The main instrumental background arises from 
$W\rightarrow e\nu$ decays, where the electron, due to 
tracking inefficiency or hard bremsstrahlung, 
is misidentified as a photon. 
This contribution is estimated from data
using a sample of isolated electrons. The same requirements
as for the {\it photon} sample are imposed, and 
the remaining number of events is scaled 
by $(1-\epsilon_{\text{trk}})/\epsilon_{\text{trk}}$, 
where $\epsilon_{\text{trk}}$
is the track reconstruction 
efficiency of ($98.6 \pm 0.1$)$\%$~\cite{diphot}.
A smaller instrumental contribution to the background is expected from
$W+\gamma$ production where the charged lepton in a leptonic $W$ boson 
decay is not detected. The kinematics of this contribution is
obtained from
$W(+\text{jets})\rightarrow \text{lepton} + \nu (+\text{jets})$
MC samples generated with {\sc pythia}, while the cross section
is taken from the MC generator based on~\cite{baur1}, which
predicts all contributions (initial state radiation,
trilinear gauge boson vertex, and final state radiation) to the
full process. 
We generate 
signal events~\cite{steve} with $M_{D} = 1.5\ {\rm TeV}$ for
$n = 2, 3, 4,5, 6,7$ and $8$.  
For different values of $M_{D}$, the cross section
scales as $1/M_{D}^{n + 2}$, leaving the kinematic spectra
unaffected for a fixed number of extra dimensions. 

\begin{table}[t]
\caption{\label{tab:summary}Data and estimated backgrounds.}
\begin{ruledtabular}
\begin{tabular}{cc}
Background&Number of expected events\\
\hline
$Z+\gamma \rightarrow \nu\overline{\nu}+\gamma$&$12.1\pm1.3$ \\
$W\rightarrow e\nu$&$3.8\pm0.3$ \\
Non-collision&$2.8\pm1.4$ \\
Misidentified jets&$2.2\pm1.5$ \\
$W+\gamma$&$1.5\pm0.2$ \\
\hline
Total Background&$22.4\pm2.5$ \\
Data&$29$ \\
\end{tabular}
\end{ruledtabular}
\end{table}

All MC events  
are passed through a detector simulation based on the 
{\sc geant}~\cite{geant}
package, and processed using the same reconstruction software
as for the data.  Additionally, we apply scale factors, with
values ranging
from $94\%$ to $98\%$, to 
account for the differences between the efficiency
determinations from data and simulation.

The main sources of systematic uncertainty are the
uncertainty in the photon 
identification efficiency ($5\%$), the uncertainty in
the total integrated luminosity ($6.1\%$), and the uncertainty
 in the signal acceptance from the PDFs ($4\%$).  

For the 
SM backgrounds estimated from MC, 
the quoted uncertainties include the uncertainty in 
the theoretical cross section, which is dominated by the uncertainty in 
the next-to-leading-order $K$ factors ($7\%$).  
For the range of $p_{T}$ in question and for the selection
requirements used in this analysis, the $K$ factors
vary around unity within this uncertainty margin~\cite{baur1,zgamma}.
The uncertainty in the width of the {\it $e/\gamma$} sample 
DCA template results in an additional systematic uncertainty
of $0.4$ events in the non-collision background estimate.

\begin{figure}[th]
\includegraphics[scale=0.43]{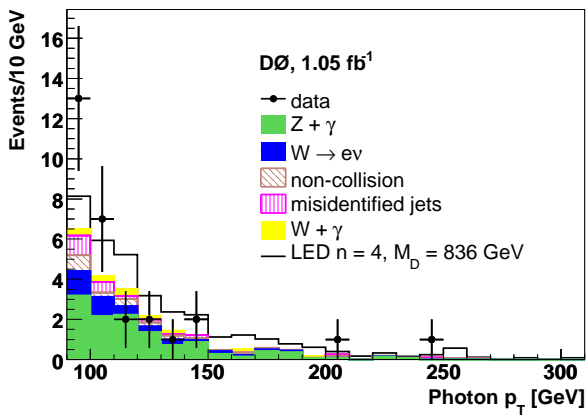}
\caption{\label{fig:pt_met} Photon $p_{T}$ distribution 
for the final candidate events (data points show statistical
uncertainties), after
all the selection requirements.  The LED signal is stacked
on top of SM backgrounds.}
\end{figure}

\begin{figure}[th]
\includegraphics[scale=0.44]{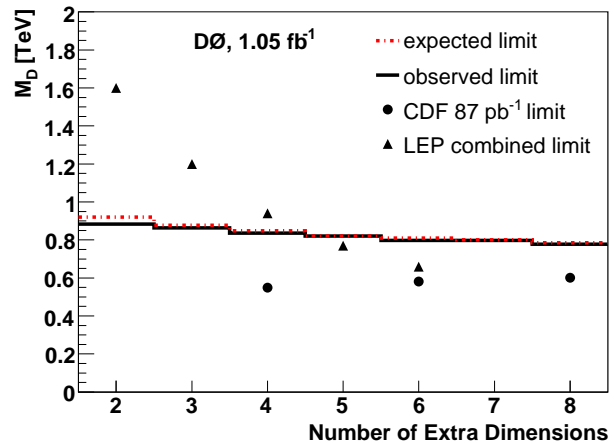}
\caption{\label{fig:limitplot} 
Expected and observed lower limits on $M_{D}$ for
LED in the $\gamma + \met$ final state.  CDF
limits with $87$~\ipb~of data~\cite{cdfrun1}, and 
the LEP combined limits~\cite{lep1} are also shown.}
\end{figure}

The final numbers of events for data and backgrounds are given in
Table~\ref{tab:summary}.  Fig.~\ref{fig:pt_met} shows the photon $p_{T}$
distribution, with the SM backgrounds stacked on top
of each other. Data and the SM expectation agree,
so we proceed to set lower limits 
for the fundamental Planck scale $M_{D}$.  
We employ the modified frequentist approach~\cite{limitc} to
set limits on the production cross section for the signal.  
This method is
based on a log-likelihood ratio test statistic and uses
the binned photon $p_{T}$ distribution. 
Assuming the leading-order theoretical cross section
for the signal, we derive the following lower limits
on $M_{D}$ at the $95\%$ C.L.:
$M_{D}> 884,~864,~836,~820,~797,~797$
and $778\ {\rm GeV}$ for $n = 2,~3,~4,~5,~6,~7$ and
$8$ extra dimensions, respectively.
Table \ref{tab:limits} and Fig.~\ref{fig:limitplot} 
summarize the results for the limit calculations.

To conclude, we have conducted a search for LED
in the $\gamma+\met$ channel, finding no evidence for
their presence.  We have set
limits on the fundamental Planck scale, significantly
improving results of previous searches.

\begin{table}[th]
\caption{\label{tab:limits}Summary of limit calculations.}
\begin{ruledtabular}
\begin{tabular}{cccc}
$n$&Signal&Observed (expected)&Observed (expected)\\ 
&efficiency&cross section&$M_{D}$ lower\\
&&limit (fb)&limit (\gev) \\

\hline
$2$&$0.49\pm 0.04$&$27.6~(23.4)$&$884~(921)$\\
$3$&$0.48\pm 0.04$&$24.5~(22.7)$&$864~(877)$\\
$4$&$0.47\pm 0.04$&$25.0~(22.8)$&$836~(848)$\\
$5$&$0.43\pm 0.04$&$25.0~(24.8)$&$820~(821)$\\
$6$&$0.50\pm 0.05$&$25.4~(22.3)$&$797~(810)$\\
$7$&$0.49\pm 0.04$&$24.0~(23.1)$&$797~(801)$\\
$8$&$0.52\pm 0.05$&$24.2~(21.9)$&$778~(786)$\\
\end{tabular}
\end{ruledtabular}
\end{table}

\input acknowledgement_paragraph_r2.tex   

\end{document}

%% file: list_of_authors_r2.tex
%
\author{V.M.~Abazov$^{36}$}
\author{B.~Abbott$^{75}$}
\author{M.~Abolins$^{65}$}
\author{B.S.~Acharya$^{29}$}
\author{M.~Adams$^{51}$}
\author{T.~Adams$^{49}$}
\author{E.~Aguilo$^{6}$}
\author{S.H.~Ahn$^{31}$}
\author{M.~Ahsan$^{59}$}
\author{G.D.~Alexeev$^{36}$}
% ...

\affiliation{\vspace{0.1 in}(The D\O\ Collaboration)\vspace{0.1 in}}
% ...
\affiliation{$^{6}$University of Alberta, Edmonton, Alberta, Canada,
                Simon Fraser University, Burnaby, British Columbia, Canada,
                York University, Toronto, Ontario, Canada, and
                McGill University, Montreal, Quebec, Canada}
% ...
\affiliation{$^{29}$Tata Institute of Fundamental Research, Mumbai, India}
% ...
\affiliation{$^{31}$Korea Detector Laboratory, Korea University, Seoul, Korea}
% ...
\affiliation{$^{36}$Joint Institute for Nuclear Research, Dubna, Russia}
% ...
\affiliation{$^{49}$Florida State University, Tallahassee, Florida 32306, USA}
% ...
\affiliation{$^{51}$University of Illinois at Chicago,
                Chicago, Illinois 60607, USA}
% ...
\affiliation{$^{59}$Kansas State University, Manhattan, Kansas 66506, USA}
% ...
\affiliation{$^{65}$Michigan State University,
                East Lansing, Michigan 48824, USA}
% ...
\affiliation{$^{75}$University of Oklahoma, Norman, Oklahoma 73019, USA}
% ...

%% file: acknowledgement_paragraph_r2.tex
%
We thank Stephen Mrenna for his help with
generating MC signal events,
the staffs at Fermilab and collaborating institutions, 
and acknowledge support from the 
DOE and NSF (USA);
CEA and CNRS/IN2P3 (France);
FASI, Rosatom and RFBR (Russia);
CNPq, FAPERJ, FAPESP and FUNDUNESP (Brazil);
DAE and DST (India);
Colciencias (Colombia);
CONACyT (Mexico);
KRF and KOSEF (Korea);
CONICET and UBACyT (Argentina);
FOM (The Netherlands);
STFC (United Kingdom);
MSMT and GACR (Czech Republic);
CRC Program, CFI, NSERC and WestGrid Project (Canada);
BMBF and DFG (Germany);
SFI (Ireland);
The Swedish Research Council (Sweden);
CAS and CNSF (China);
and the
Alexander von Humboldt Foundation.

%% file: led.bbl
\begin{thebibliography}{999}
\bibitem{pseudo} Pseudorapidity is defined as $-\ln[\tan(\theta/2)]$, where
$\theta$ is the angle between the particle and the proton beam direction.
\end{thebibliography}
